\def\lesssim{\mathrel{\mathchoice {\vcenter{\offinterlineskip\halign{\hfil $\displaystyle##$\hfil\cr<\cr\sim\cr}}} {\vcenter{\offinterlineskip\halign{\hfil$\textstyle##$\hfil\cr <\cr\sim\cr}}} {\vcenter{\offinterlineskip\halign{\hfil$\scriptstyle##$\hfil\cr <\cr\sim\cr}}} {\vcenter{\offinterlineskip\halign{\hfil$\scriptscriptstyle##$\hfil\cr <\cr\sim\cr}}}}}
\def\grtsim{\mathrel{\mathchoice {\vcenter{\offinterlineskip\halign{\hfil $\displaystyle##$\hfil\cr>\cr\sim\cr}}} {\vcenter{\offinterlineskip\halign{\hfil$\textstyle##$\hfil\cr >\cr\sim\cr}}} {\vcenter{\offinterlineskip\halign{\hfil$\scriptstyle##$\hfil\cr >\cr\sim\cr}}} {\vcenter{\offinterlineskip\halign{\hfil$\scriptscriptstyle##$\hfil\cr >\cr\sim\cr}}}}}

\documentclass{cjaa}                    

\usepackage{graphicx}                   
\input{epsf.sty}                        
\input{psfig.sty}                       

\begin{document}

   \title{Gamma-Ray Bursts: explained my way}

   \author{Wolfgang Kundt}
      \institute{Institut f\"ur Astrophysik der Universit\"at, Auf dem H\"ugel 71, D-53121 Bonn, Germany\\ 
      \email{wkundt@astro.uni-bonn.de}
    }

   \offprints{W. Kundt}                  

  \date{Received~2003; accepted~2003}

\abstract{
Ordered historically, I shall update my earlier conviction that a consistent interpretation of all the non-terrestrial gamma-ray bursts can be obtained in terms of nearby Galactic neutron stars, at distances $d$ within 10 $\lesssim d$/pc $\lesssim$ 500.
}
   
   \authorrunning{W. Kundt}            
   \titlerunning{Gamma-ray bursts}    

   \maketitle

\section{Era of Convergence}           
\label{sect:convergence}

The $\gamma$-ray bursts made their scientific appearance at the 1974 Texas Symposium on Relativistic Astrophysics when they had just been declassified by American militaries, and were introduced to the community by Malvin Ruderman. In his superb presentation, Ruderman (1975)
offered more proposed classes of sources than detected bursts, viz. Expanding Supernova Shocks, Neutron Star Formation, Glitches, Neutron Stars in Close Binaries, Black Holes in Binaries, Novae, White Holes, Flares on \{``Normal" Stars, Flare Stars, White Dwarfs, Neutron
Stars\}, or in Close Binaries, Nuclear Explosions on White Dwarfs, Comets on Neutron Stars, Jupiter, Anti-matter on Conventional Stars, Magnetic Instabilities in the Solar Wind, Relativistic Dust, Vacuum Polarization Instabilities near Rotating Charged Black Holes, Instabilities in Pulsar
Magnetospheres, "Ghouls", and various combinations thereof. In his Conclusions, he admitted that he had no firm prediction yet, but favoured two horses the first of which is still up in the race today: black-hole accretion.

Some convergence of interpretation took place during the succeeding ten years, whereby the energetics, time scales, hardnesses, and spectra asked for the compactness, and deep potential well of a neutron star's surface, the richness of different lightcurves (composed of `FREDs') asked for a
composite process, of strong non-uniformity, and the occurrence rates were consistent with the population of nearby, old Galactic neutron stars (Mazets et al 1980,1981, with their redshifted annihilation lines, and cyclotron lines; Verter 1982, Epstein 1986, Higdon \& Lingenfelter 1990, Harding 1991, Lingenfelter \& Higdon 1992, Fenimore et al 1992, Ryan et al 1994). 

More in detail, (i) the Eddington power of $10^{38.5}$erg/s of a neutron-star source asks for source distances $\lesssim \gamma$ Kpc, ($\gamma$ = Lorentz factor of beaming), (ii) temporal finestructure down to $\grtsim 10^{-3.7}$s asks for source sizes of $\lesssim$ tens of Km, (iii) spectra ranging out to MeV energies ask for either transrelativistic potential wells, or
transient electric voltages exceeding MV, (iv) the richness of different lightcurves is reminiscent of Jupiter's accretion of comet Shoemaker-Levy (in 1993) -- torn into a string of clumps by tidal forces -- and (v) the on average observed four bursts per day hitting Earth carry the power expected from spasmodic interstellar accretion onto old Galactic neutron stars, some $10^{-17} M_{\odot}$ per year and neutron star, (cf. Kundt 2001).

By the mid 90s, it had become clear that the detected GRBs form a 3-humped distribution: Largest is the subset of long bursts, of duration between $1$ and $10^3$s, peaking at 30 s; perhaps 40\% of all bursts are short bursts, of duration between $10^{-2}$s and 3 s, with peak at 0.3 s; and
there is a disjoint, tiny subset, $\grtsim 1\%$ of all, of duration between $\lesssim 1$ msec and 5 msec, which is harder by one in spectral index, and may (equally) extend up to MeV energies, perhaps even much higher ($\lesssim 10^2$ MeV). This third subset is of terrestrial origin, emitted
during mesospheric discharges, in connection with red, horizontal, ringlike `elves' at the ionosphere, blood-red, vertical, fibrous, mesospheric `sprites', blue, inclined, stratospheric `jets', and `gigantic jets' which span the whole range of heights, from the blue jets up to the red sprites
all the way to the ionosphere, (at some 90 Km: Mende et al 1997, Su et al 2003). The terrestrial GRBs show us that (even) charged, low-density plasmas can be the sources, on scales of tens of Km, though at much lower (intrinsic) power than their extraterrestrial cousins. 

It is not clear yet why the extraterrestrial GRBs form a 2-humped set. Afterglows are found exclusively for long bursts, yet only in about half of all cases. In the Galactic neutron-star model, afterglows result as reflection nebulae, by the light-echo effect: The burst, generated at the star's
surface by impacting clumps (`blades') of interstellar matter, illuminates its environs which radiate via fluorescence, with a spectral power ($\nu S_{\nu}$) that tends to rise as a flattish power law, from radio to X-ray frequencies. The different observed afterglow lightcurves result from different
circumstellar mass distributions. They are much brighter for the long bursts than for the short bursts because ionized matter is centrifugally ejected from the star's surface during each splash, via its corotating magnetosphere, and takes about a time of 1/$\Omega \approx$ 1 s to reach light-cylinder distances, ($\Omega$ = spin angular frequency), where it intersects all subsequent
radiation from the surface. Because it is so near to the source, this freshly ejected matter dominates the early afterglow. Its transrelativistic speed when crossing the light cylinder, at an angle near $45^0$ w.r.t. the radial direction, causes a (transverse) Doppler redshift of its
absorption which is routinely translated into a cosmic distance, even though SS 433 shows us that nearby neutron stars can do it as well. No surprise that the X-ray afterglows tend to show Fe and Mg in their early (redshifted) absorption spectra, in particular strong Fe K-edge absorption (Amati et al 2000), scraped off a neutron star's surface.

This alternative interpretation of GRB afterglows has been proposed in (Kundt 2001a) where I estimated a transient wind number rate of $\dot N = 2 \pi c h N_H = 10^{40} s^{-1} r_{10.5}$, corresponding to a transient mass-loss rate of $\dot M = 10^{-10} M_{\odot}/yr$, in which $r$ is the radius of the inner edge of an escaping wind column, of H-column density $N_H$, height $h$, and $r_{10.5} := r/10^{10.5}$cm. Unfortunately, the original formula contains two (minor) printing errors.  

In this alternative interpretation of GRB afterglows, as Galactic light echos, I have assumed that the SGRs are representative of the whole class, being the nearest -- and hence brightest -- among them, at distances of $\lesssim 50$ pc, (for which we do not only see their `ordinary' bursts but also their much more frequent, softer, and much dimmer repetitions, at a $10^3$-times lower power). For SGR 0525-66, 1806-20, 1907+09, and 1627-41, we know their spin periods, P/s = 8.0, 7.47, 5.17, and 6.4, yielding afterglow delay times of  P/2$\pi$ s = 1.3, 1.2, 0.82, and 1.0 respectively, as anticipated in the preceding paragraph. 

With this interpretation, I have assumed that the most famous SGR, GRB790305, only projects onto SNR N 49 in the LMC but, in reality, lies at the inner edge of the GRB source distribution (in the Galaxy), at a distance of some $50$ pc, as has been convincingly argued by Zdiarski (1984), also Bisnovatyi-Kogan \& Chechetkin (1981). Similarly, none of the other SGRs has a well-determined distance: Their embedding synchrotron nebulae are likely pulsar nebulae, not SNRs, whose intrinsic luminosities are much fainter than assumed in the distance estimates (Kundt \& Chang 1994).    

\section{Era of Confusion}
\label{sect:confusion}

A first, seeming blow to the nearby-Galactic neutron-star model came from BATSE statistics, with the increasing isotropy of arrival directions of the bursts by the end of 1991. In those days, stimulated by the evidence, I noticed that the accretion disks of old Galactic neutron stars are expected to
have their angular-momentum vectors (roughly) perpendicular to that of the Milky-Way disk, because assembled by the stars from interstellar clouds during oscillatory motion up and down through the disk, and that their accreting chunks (blades) would radiate preferentially in the assembled disk planes, like cooling sparks
emitted by a grindstone. As a result, the probability of being in the beam of a Milky-Way GRBer decreases roughly by a factor of ten with decreasing galactic latitude, starting with unity in polar direction, and we expect a first-order compensation between column length and beaming  probability, i.e. an almost isotropic distribution of bursts, as stated above (in the abstract). 

Kundt \& Chang (1993) played this game both analytically and numerically, and were able to reproduce the measurements for plausible ranges of three fit parameters (angles). Our predictions have remained valid for the improved statistics analysed by Belli (1997), by Bal\'{a}zs et al (1997), and by
Pendleton et al (1997), all three of whom meet with problems in the cosmological interpretation. (Belli finds anisotropic distributions for the subclasses of short-hard bursts and long-soft ones, and Bal\'{a}zs et al see the local spiral arm in their data, as a quadrupole component). The local interpretation rescues our awareness of the dead-pulsar population, whose accretion power must not have gone unnoticed throughout 35 years.  

A second, seeming blow to the nearby-Galactic neutron-star model came in 1997 with Beppo-SAX, whose X-ray pointing accuracy of 3' allowed a large number of afterglow discoveries, and a hunt for host galaxies. Clearly, with the advent of the large-aperture Keck telescope and the HST,
time was just right to look deeper into the Cosmos, and discover luminous patches in projection onto the burst locations which would not have been recognisable in earlier decades. The afterglows at X-rays revealed redshifts which were interpreted as distances (rather than as Doppler shifts from relativistic
ejecta), and the optical light echos were interpreted as distant host galaxies. Little
attention was paid to the absence of long-distance travel signatures in the burst lightcurves (Mitrofanov 1996), to the ``no-host-galaxy dilemma" of Schaefer et al (1997), or to the large spread in redshifts of the
afterglows (from z = 0.0083 to z = 4.50), inconsistent with the thin-shell distribution read off the BATSE catalogue. A large number of further inconsistencies in the cosmological interpretation of the bursts have been collected by Lamb (1999), and more recently by Bisnovatyi-Kogan (2003).

If the burst sources had distances of order 10 Gpc rather than 0.1 Kpc, their powers would be some $10^{16}$ times larger than in the local interpretation -- which latter deals with sub-Eddington sources -- challenging theorists to invent new and exotic classes of cataclysmic phenomena. Invented were magnetars, fireballs, cannon balls, dancing jets, hypernovae, collapsars, and more exotic scenarios. `Jets' were invoked
as though these phenomena had anything in common with the hundreds of well-mapped local jet sources, ignoring that the latter are quasi-stationary, ramming their vacuum channels on the time scale of years, whereas the bursts depend on phenomena that form
and evolve within fractions of a second. It's not only the huge inferred energies -- at $\gamma$-rays -- that matter but even more so the short time scales on which they would have to be liberated, i.e. it's the even much huger powers (at $\gamma$-rays (!)).

As I mistrust all the published host-galaxy identifications, I asked Eli Waxman (2003), the reviewer in Nature of the strongly polarized (!) burst GRB021206, to tell me the best case of a host-galaxy identification. He named GRB970508, and directed my attention to its identification papers by
Bloom et al (1998) and by Fruchter et al (2000). The host is peculiar for its elongated shape, blue spectrum, and for housing the burst within $\lesssim 70$ pc of its center, without any extinction signature in the afterglow spectra. Its spectrum contains (only) two emission lines, `identified' as
[Ne III]3869 and [O II]3727, whose redshifts agree with those of the afterglow. But Allen's table of spectral lines tells me that the line [Ne III]3342 should be 16 times stronger than the identified line; it is absent. Perhaps, the two lines should be re-interpreted as coming from nearby [Fe XV]
and [Ni XV], and we see a glowing reflection nebula. You will understand that I mistrust all the 20 `host galaxies' discussed by Bloom et al (20002) and by Masetti et al (2003), which contain GRB970508 as their best case. 

In the meantime, Manfred Pakull has informed me that the suspicion expressed in the preceding paragraph was not fully justified: [Ne III]3342 is a nebular line whereas [Ne III]3869 is an
auroral line. Their intensity ratio can vary by a factor of $10^3$, depending on the ratio of
collisional over radiative excitations, cf. Lang (1980, p. 107).

The confusion about GRB sources was further enhanced when similarities were discovered between (piecewise exponential) SN lightcurves and a number of (piecewise power-law) optical afterglows, and even between segments of their spectra, cf. Price et al (2003). Statistically, near
coincidences in the sky of SNe and GRBs are practically ruled out, because both phenomena are rare. The SN piston takes hours to reach the surface of the progenitor (supergiant) star, whereas a long GRB takes typically 30 s -- how can the two phenomena be related ? The variety of
broadband, densely sampled afterglow lightcurves is still rapidly increasing, cf. Pandey et al (2003), as is expected for light echos from different circumstellar environments plus different transient (relativistic) winds, but hardly from a uniform class of cosmic super events.

Let me end this appeal for the nearby-Galactic neutron-star interpretation by listing four further difficulties for the cosmological interpretation. They are: (i) the recently observed high polarization of GRB021206 (at $\gamma$-rays, $(80 \pm 20)\%$, Coburn \& Boggs 2003), reminiscent of the
strong surface magnetic fields of neutron stars; (ii) the `X-ray transients' emphasized by John Heise (2003), also Piro (2003): soft equivalents of the GRBs at $\grtsim$ 1/3 their rate, with otherwise quite similar properties; (iii) the occasionally very hard, and long-lasting spectra of some 10 bursts, extending up in energy to 0.2
GeV, to 20 GeV, or even to $\grtsim$ 10 TeV (!), Kundt (2001a), Bisnovatyi-Kogan (2003), even with $\nu S_{\nu}$ rising all the way to the upper end of the recorded spectrum, at 0.2 GeV for GRB941017, Gonz\'{a}lez et al (2003); such spectra are familiar from pulsars; and (iv) the X-ray
precursors of GRBs, by $\lesssim 10^{3.7}$s, highlighted by Daniele Fargion (2003). 

Each of these facts tells us that no convincing cosmological interpretation is in sight.  

\section{Conclusions}
\label{sect:conclusion}

New evidence is coming in, with every well-sampled burst and/or afterglow, that the cosmological interpretation of the GRBs is untenable, with its excess factor of $10^{16}$ in burst power over local sources. It would never have boomed, had the coming generation been brought up to discriminate between proposed alternatives rather than to strengthen a growing consensus.
Nearby Galactic neutron stars satisfy all the constraints.

\begin{acknowledgements}
Friendly thoughts go to those participants -- including a subset of speakers of Concluding Remarks -- who attended this talk.
\end{acknowledgements}

\label{lastpage}

\end{document}